\documentclass{aa}
\usepackage{graphicx}
\usepackage{natbib}  
\usepackage{txfonts}

\begin{document}
   \title{High\--resolution CRIRES spectra of Terzan~1: a metal\--poor 
   globular cluster toward the inner bulge\thanks{Based on data taken at the 
   ESO/VLT Telescope, within the observing program 093.D-0179(A)}.}

   \author{E. Valenti \inst{1}
	  \and L. Origlia\inst{2}
	  \and A. Mucciarelli \inst{3} 
	\and R.M. Rich \inst{4}
          }

\institute{European Southern Observatory, Karl-Schwarzschild-Str. 2, 
D-85748 Garching bei Muenchen, Germany
         \and
             INAF - Osservatorio Astronomico di Bologna,
             Via Ranzani 1, I-40127 Bologna, Italy
	     \email{livia.origlia@oabo.inaf.it}
         \and
             University of Bologna, Physics \& Astronomy Dept.,
             Viale Berti Pichat 6-2, I-40127 Bologna, Italy
         \and
	     Physics and Astronomy Bldg, 430 Portola Plaza Box 951547 Department 
	     of Physics and Astronomy, 
             University of California at Los Angeles, Los Angeles, CA 90095-1547, US
             }

\authorrunning{Valenti et al.}
\titlerunning{CRIRES spectra of Terzan~1}

   \date{Received .... ; accepted ...}

 
  \abstract
   {}
   {Containing the oldest stars in the Galaxy, globular clusters toward
   the bulge can be used to trace its dynamical and chemical evolution.
   In the bulge direction, there are $\sim$50 clusters, but only about 20\% have been subject 
   of high\--resolution spectroscopic investigations. So far, the sample observed at 
   high resolution spans a moderate\--to\--high metallicity regime. In this sample,
   however, very few are located in the innermost region ($R_{GC}\leq$1.5\,Kpc and $|l, b|\leq5^{\circ}$).
   To constrain the chemical evolution enrichment of the innermost region of Galaxy,
   accurate abundances and abundance patterns of key elements based on high\--resolution
   spectroscopy are necessary.
   Here we present the results we obtained for Terzan\,1, a metal\--poor cluster located in 
   the innermost bulge region.
   }
   {Using the near\--infrared spectrograph CRIRES at ESO/VLT, we obtained high\--resolution 
   (R$\approx$50,000) H\--band
   spectra of 16 bright giant stars in the innermost region ($r\leq 60"$) of Terzan~1. 
   Full spectral synthesis techniques and equivalent width measurements of selected lines, isolated
   and free of significant blending and/or contamination by telluric lines, allowed
   accurate chemical abundances and radial velocities to be derived. 
   }
   {Fifteen out of 16 observed stars are likely cluster members, with an average heliocentric 
   radial velocity of +57$\pm$1.8\,km/s and mean iron abundance of [Fe/H]=--1.26$\pm$0.03\,dex. 
   For these stars we measured some [$\alpha$/Fe] abundance ratios, finding average values of
   [O/Fe]=+0.39$\pm$0.02\,dex, [Mg/Fe]=+0.42$\pm$0.02\,dex, [Si/Fe]=+0.31$\pm$0.04\,dex, and 
   [Ti/Fe]=+0.15$\pm$0.04\,dex
   }
   {The $\alpha$ enhancement ($\approx +0.4$\,dex) found in the observed giant stars of Terzan~1
   is consistent with previous measurements on other, more metal\--rich bulge clusters, which 
   suggests a rapid chemical enrichment.}


   \maketitle
%

\section{Introduction}

Observational and theoretical studies have long shown that Galactic globular clusters 
(GCs) contain the oldest stars in the Galaxy. 
They represent an important tracer of the underlying population, therefore deriving accurate 
abundances and abundance patterns in GCs represents a crucial test for tracing the early dynamical 
and chemical evolution of the Galaxy.
The elemental abundance distributions and the abundance ratio of certain critical elements,
such as Fe-peak, CNO, and $\alpha$-elements (i.e., those synthesized from $\alpha$ particles, such as O, 
Ne, Mg, Si, Ti, Ca and S) are particularly suitable for this purpose. 
Indeed, these elements are synthesized in stars of different masses, hence released into
the interstellar medium on different timescales. 
A particularly useful abundance ratio is [$\alpha$/Fe]. 
Thanks to the time delay in the bulk of Fe production relative to $\alpha$-elements, the [$\alpha$/Fe] 
abundance ratio can be efficiently used as a cosmic clock \citep[see, e.g.,][]{mw97}. 

An overall [$\alpha$/Fe] enhancement with respect to the solar value has been well established in the 
Galactic halo GCs population for several years \citep[see][for a general review]{gra04},
indicating a major enrichment by type II\,SNe on a short timescale.
On the other hand, accurate abundance patterns are only available for some GCs in the bulge 
direction. 
Within a total population of about 50 GCs located in the bulge \citep[][2010 compilation]{harris}, 
only about 20\% have been subject 
to spectroscopic investigation at moderately high spectral resolution (R$\ge$20,000) for detailed 
chemical abundance analysis. 
Our team has contributed significantly to this subject by measuring chemical abundances in cool 
giants that are members of 12 bulge GCs \citep{ori02,ori04,ori05a,ori05b,ori08,val11} by using H-band 
high-resolution spectra obtained at KeckII with NIRSPEC \citep{ml98}. 
Chemical abundances from high\--resolution optical spectroscopy in seven other bulge GCs have been obtained 
by \citet{bar99,bar06,bar07,bar14,cohen99,eug01,eug07,zoc04,albr06,gra06}.


  \begin{figure*}
  \centering
  \includegraphics[width=9cm , height=9.5cm]{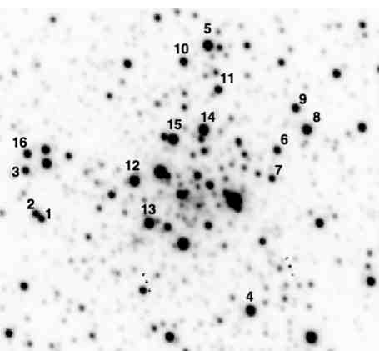}
  \includegraphics[width=9cm , height=9.5cm]{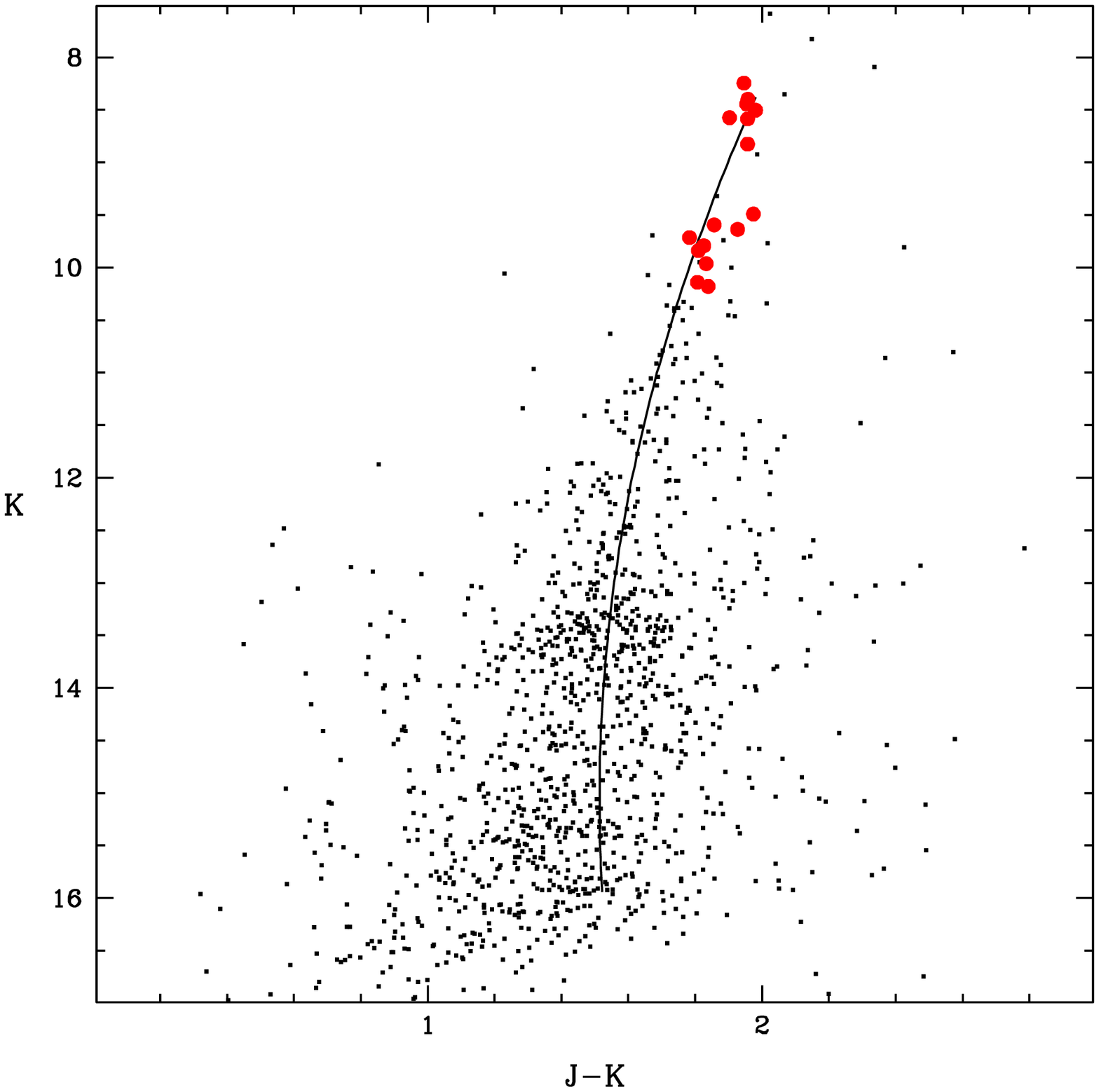}
   \caption{Left panel: SofI K\--band image of the core region of Ter~1. The field of view shown 
   in the map is $\sim 1'\times 1'$, north is up and east on the left. 
   The observed stars are numbered (cf. Table~\ref{tab1}). Right panel: Observed K, (J-K) CMD, and 
   mean RGB ridgeline of Ter~1 in the central $\sim$ 1 arcmin in radius from \citet{val10}. 
   Large red 
   filled points denote the stars observed spectroscopically.}
              \label{map-cmd}
    \end{figure*}

From the measured Fe, O, Ti, Si, Mg, and Ca abundances, we found that the [$\alpha$/Fe] ratios are 
enhanced by a factor between 2 and 3 over the whole range of metallicity spanned by the observed 
clusters. All these measurements indicate that the bulk of the bulge GC population probably formed 
from a gas that is mainly enriched by type II\,SNe and on a short timescale, before substantial 
contribution of type Ia\,SNe took place. 

With the only exception of Terzan~4 \citep{ori04}, which has a metallicity [Fe/H]=--1.6~dex, 
all the bulge GCs observed so far at high spectral resolution have metallicity 
[Fe/H]$\ge$--1.0~dex. In this sample observed at high\--resolution, there are very few clusters
located in the very innermost bulge region (i.e., $R_{GC}\leq$1.5\,Kpc and $|l , b|\leq5^{\circ}$).
 
With the aim of understanding the chemical composition of the innermost stellar populations 
in the Galactic bulge, we used CRIRES at the VLT \citep{crires} to measure the chemical 
composition of a few GCs located toward the center of the bulge.
Here we present the results for Terzan~1. 

Terzan~1 is a GC located close to the plane, with somewhat controversial reddening and distance 
estimates. An optical photometric study with HST-WFPC2 by \citet{ort99} suggested E(B--V) = 2.48 
and (m--M)$_0$=13.58.
A near-IR study by \citet{val10} suggested a lower reddening E(B--V)= 1.99 and a larger distance 
modulus (m--M)$_0$=14.13.
However, when taking the different $\Delta$E(B--V)=0.5 into account, the two distance estimates agree within 
the errors ($\sim$0.2 mag).
From the morphology (i.e., color, magnitude, and slope) of the cluster's red giant branch (RGB) in 
the (K,J--K) color-magnitude diagram (CMD), we derived a photometric metallicity [Fe/H]=-1.1 dex 
\citep{val10}.
Low-resolution optical spectroscopy of 11 giants that are likely members of 
Terzan~1 which was obtained by \citet[][hereafter I02]{idi02}, suggests a 
[Fe/H]$\approx$--1.3 and some [Mg/Fe] enhancement with respect to the solar ratio.

\section{Observations and data reduction}
\label{obs}

To minimize the risk of contamination from bulge field stars, we selected 16 bright giants
($8 \le K \le 10.5$) using the near-IR CMD by \citet{val10} in the innermost region (at r$\leq$60") 
and along the mean RGB ridgeline (see  Fig.~\ref{map-cmd}). Table~\ref{tab1} lists the near\--IR
magnitude and coordinates of the selected stars, whereas Fig.~\ref{map-cmd} shows 
their spatial distribution within the cluster.

Observations were executed in service mode, during several nights in April and June 2014 
and under an average seeing of $\approx$ 1" .
We used CRIRES in non\--AO mode and selected a slit width of 0.4" providing a spectral resolution 
R$\approx$50,000 
and two different gratings in the H band, namely 1550.8\,nm/order\#36 and 1598.0\,nm/order\#36.
However, because of the known vignetting and noise problems affecting the four detectors, only 
portions of the spectra in the ranges:
1.547-1.553 $\mu m$, 1.557-1.635 $\mu m$, 1.575-1.581 $\mu m$, 
1.585-1.590 $\mu m$, and 1.595-1.601 $\mu m$ were actually usable.    

The integration time, on source and per setting, was 20 min.
The data was reduced by using the CRIRES 
pipeline\footnote{http://www.eso.org/sci/software/pipelines}, which performs sky
subtraction using the pair of spectra obtained when the object is nodded along the slit and  
flat\--fielding correction. Wavelength calibration was performed using 
a ThAr lamp and a second\--order polynomial solution. Because of the gaps within the four CRIRES detectors
and the relatively narrow wavelength range covered by each setting, the derived wavelength solution using 
the ThAr lamp strongly depends upon the number and distribution of available lines within a given detector. 
Therefore, to check and refine the pipeline wavelength solutions, we cross\--correlated (using the
IRAF task {\it fxcor})  the spectra of an O star observed for telluric correction
before or after the targets, with a high\--esolution spectrum of the Earth's telluric features
retrieved from the ESO web 
page\footnote{http://www.eso.org/sci/facilities/paranal/decommissioned/
isaac/tools/spectroscopic\_standards.html}. The cross\--correlation between the template and the O\--type star 
spectra was performed over a region around 1.577$\mu$m where several telluric features are present.

\begin{table}
\caption{Stars toward Terzan~1 observed with CRIRES and selected from \citet{val10} catalog.}
\label{tab1}     
\centering                          
\begin{tabular}{cccccc}  
\hline\hline
ID & J & H & K & RA (J2000) & DEC (J2000) \\
\hline
%
1 &   11.947 &   10.564 &   10.140  &   263.9564516  &    -30.4829633 \\  
2 &   11.796 &   10.392 &    9.963  &   263.9568131  &    -30.4826890 \\  
3 &   11.648 &   10.250 &    9.838  &   263.9574730  &    -30.4803350 \\  
4 &   10.542 &    9.088 &    8.585  &   263.9430320  &    -30.4880850 \\  
5 &   10.478 &    -     &    8.575  &   263.9457667  &    -30.4733797 \\  
6 &   11.618 &   10.237 &    9.792  &   263.9413450  &    -30.4792040 \\  
7 &   12.018 &   10.618 &   10.179  &   263.9416433  &    -30.4807381 \\  
8 &   10.783 &    9.294 &    8.826  &   263.9394140  &    -30.4780540 \\  
9 &   11.565 &   10.113 &    9.638  &   263.9401400  &    -30.4768680 \\  
10 &  11.450 &   10.028 &    9.593  &   263.9473110  &    -30.4742850 \\  
11 &  11.498 &   10.123 &    9.715  &   263.9451520  &    -30.4758380 \\  
12 &  10.398 &    8.960 &    8.444  &   263.9504970  &    -30.4808830 \\  
13 &  10.485 &    9.020 &    8.504  &   263.9495090  &    -30.4832320 \\  
14 &  10.356 &    8.913 &    8.398  &   263.9460670  &    -30.4781060 \\  
15 &  10.191 &    8.688 &    8.245  &   263.9480840  &    -30.4785630 \\  
16 &  11.466 &   10.045 &    9.492  &   263.9573380  &    -30.4793950 \\  
\hline
\hline
\end{tabular}
\end{table}

  \begin{figure*}
  \centering
  \includegraphics[width=\hsize]{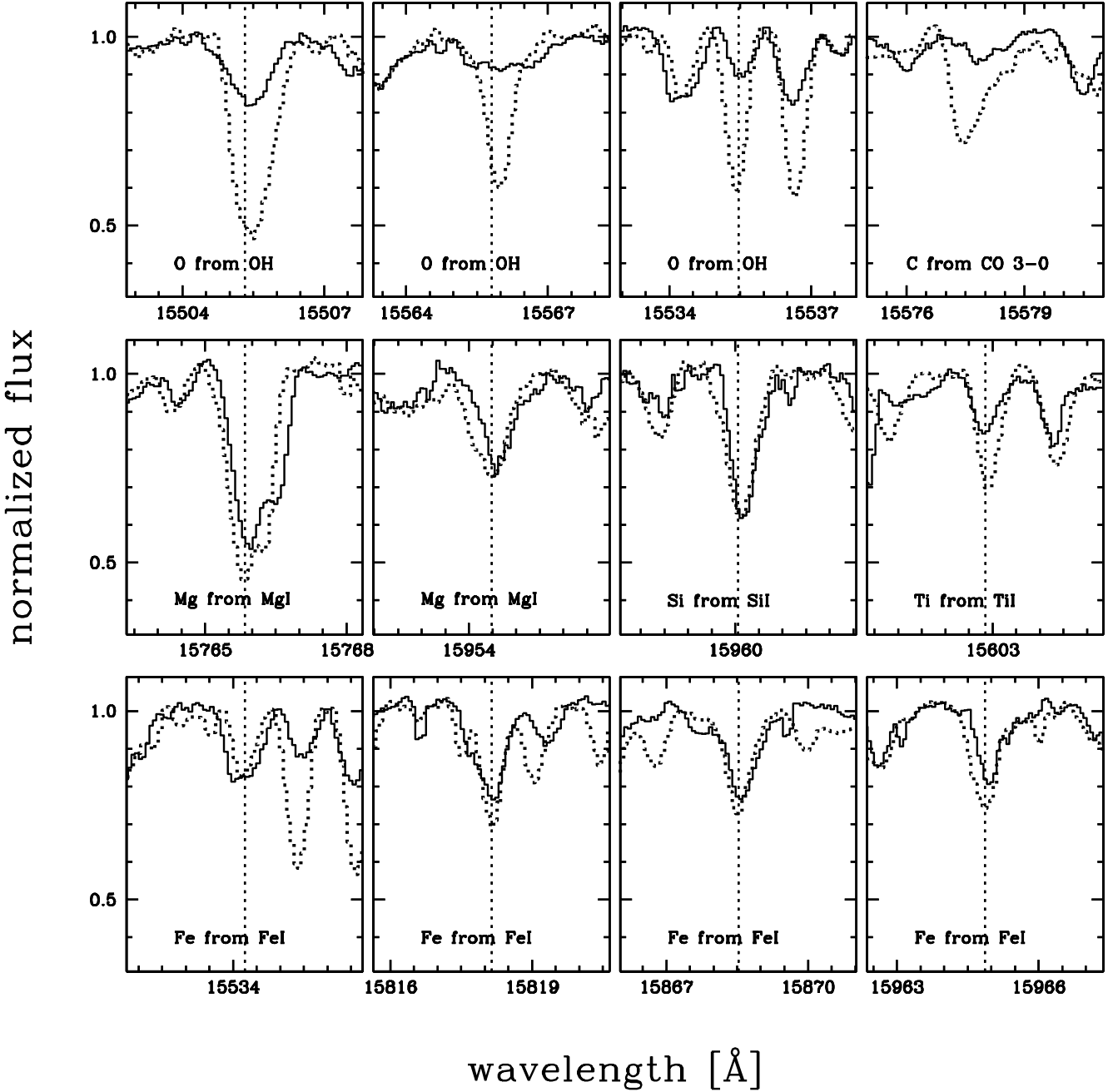}
   \caption{CRIRES spectra centered on some lines of interest of two giant stars 
   (\#11, continous line and \#12, dotted line) with very similar abundances but very different 
   temepratures. The coolest giant (\#12 at T$_{eff}$=3600~K) has much deeper molecular OH and CO 
   lines than the warmest one (\#11 at T$_{eff}$=4250~K).}
              \label{spectra}
    \end{figure*}

\section{Chemical abundance analysis}

To measure the chemical abundances of the RGB stars in Terzan~1 from the CRIRES spectra, 
we performed the same analysis as in our previous works on bulge GC and field 
giants \citep[see][and references therein]{val11,ric12}.

We made use of full spectral synthesis techniques 
and the equivalent width measurements of selected lines, which were 
sufficiently isolated, free of significant blending and/or 
contamination by telluric absorption.
Telluric absorptions were carefully 
checked on an almost featureless O-star spectrum.

To compute suitable synthetic spectra
and model the observed giant stars, 
we used an updated version \citep{ori02} of the code    
first described in \citet{ori93}.
The code uses the LTE approximation, which is based
on the molecular blanketed model atmospheres of
\citet{jbk80} at temperatures $\le $4000~K
and the ATLAS9 models for temperatures above 4000~K. It
includes thousands of near IR atomic transitions 
from the Kurucz database\footnote{http://www.cfa.harvard.edu/amp/ampdata/kurucz23/
sekur.html},
from \citet{bie73}, and from \citet{mel99}. 
Molecular data are from our compilations \citep[][ and subsequent updates]{ori93} 
and from B. Plez (private communication).
The reference Solar abundances are from \citet{gre98}.

An initial guess of the temperature and gravity of the observed stars has been derived from 
their near-IR photometry,
while for the microturbulence velocity we adopted an average value of 2.0 km/s
\citep[see also][]{ori97}.
These photometric estimates of the stellar parameters were used
as input to produce a grid of model spectra that span a wide range in terms of
abundances and abundance patterns, while keeping the stellar parameters (temparature and
gravity) around the photometric values.
The model that reproduces the overall observed spectrum and
the equivalent widths of selected lines better was chosen as the best fit model.
Equivalent widths were computed by Gaussian fitting
the line profiles, with an overall uncertainty $\le$10\%.

We have been able to reliably measure a handful of Fe and OH lines, 
and (depending on the signal-to-noise of the spectra) one or two Ti, Si, and Mg lines. 
Equivalent width measurements of these lines 
were used to get abundances of Fe, O, and other three (Mg, Si, and Ti) alpha-elements.  
Carbon abundances are mostly determined from the 3-0 $^{12}$CO bandhead 
by means of full spectral synthesis. 
A few, quite faint CN lines are also present in the spectra of the observed stars but the 
signal-to-noise is not adequate to get reliable nitrogen abundances.

For each observed star, Table~\ref{tab2} lists
the final adopted   
photospheric parameters, the measured radial velocity, and chemical 
abundances. In particular, the tabulated star temperature and gravity values are photometric,
but spectroscopically fine\--tuned to get simultaneous spectral fitting of the CO bandhead, 
the OH molecular lines,
and the atomic lines observed in the CRIRES spectra.
The impact of using slightly different assumptions for the stellar parameters on the derived 
abundances is discussed in Sect.~\ref{error}.
However, it is worth mentioning that since the CO and OH molecular line profiles are very
 sensitive to
effective temperature, gravity, and microturbulence variations,
they reliably constrain the values of the phtospheric  parameters, therefore
significantly reducing their initial range of variation and
ensuring a good self-consistency of the overall spectral
synthesis procedure \citep{ori02,ori04}.
As a clear example, 
Fig.~\ref{spectra} shows the CRIRES spectra centered on some lines of interest of two 
giant stars (\#11 and \#12) with very similar abundances but very different temperatures.
The coolest giant (\#12 at T$_{eff}$=3600~K) has much deeper molecular OH and CO lines 
than the warmest one (\#11 at T$_{eff}$=4250~K).

  \begin{figure*}
  \centering
  \includegraphics[width=\hsize]{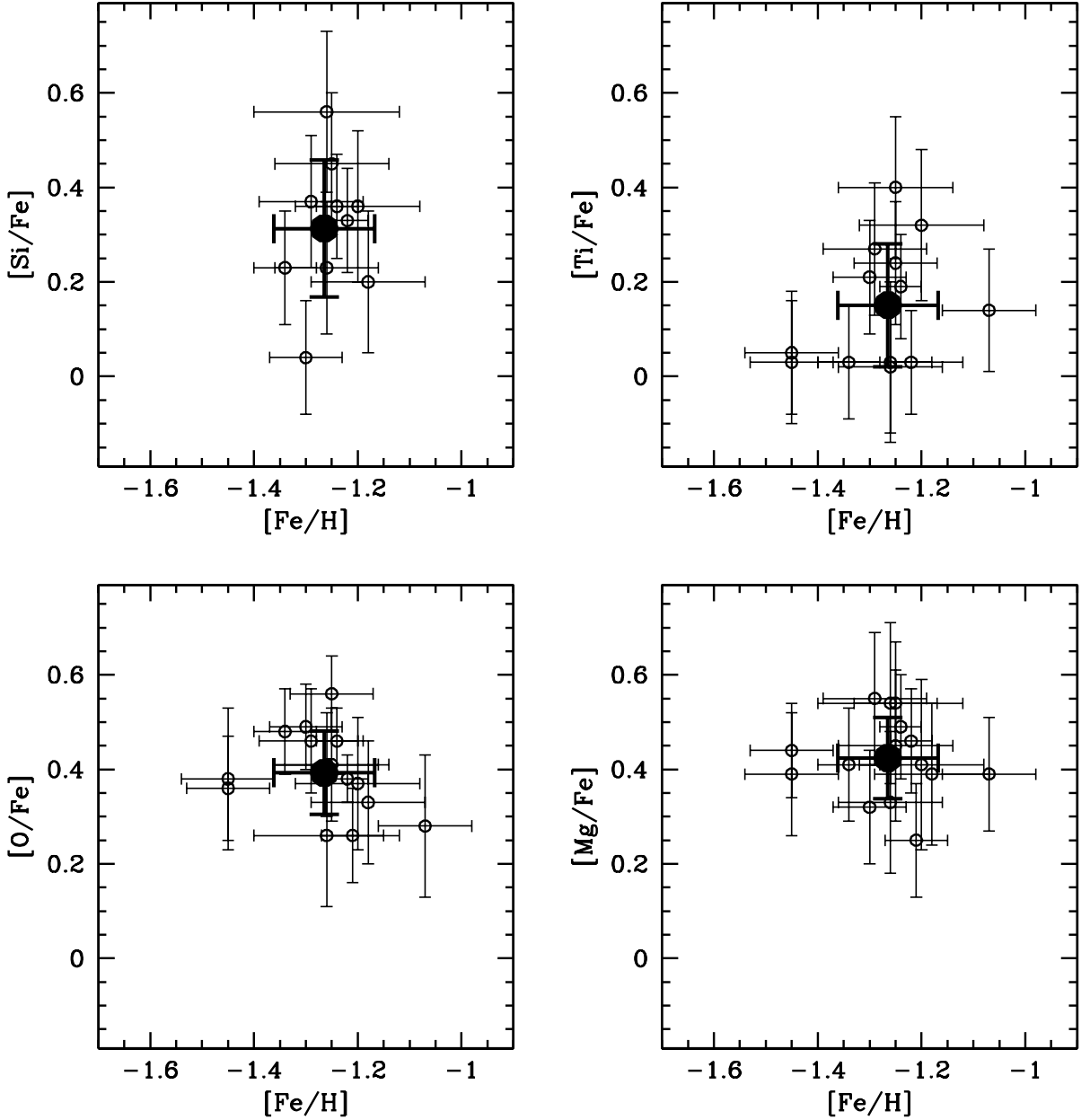}
   \caption{[$\alpha$/Fe] abundance ratios as a function of [Fe/H] for the 15 stars likely cluster members.
The big dots indicate the average values and 1$\sigma$ dispersions.}
              \label{alpha}
    \end{figure*}

\subsection{Error Budget}
\label{error}

We can quantify random and systematic errors in the measurement of the equivalent widths and in the derived 
chemical abundances as follows.
The typical random error of the measured line equivalent widths is 10\%, arising mostly 
from a $\pm $2\% uncertainty in the placement of the pseudo-continuum, as estimated 
by overlapping the synthetic and the observed spectra. 
Such random uncertainties in the line equivalent width measurements correspond to abundance variations 
ranging from a few hundredths to 1 tenth of a dex.
This $\le $0.1 dex error is somewhat lower than the typical 1$\sigma $ scatter in the derived abundances 
from different lines, which normally ranges between 0.1 and 0.2 dex.
The errors quoted in Table~\ref{tab2} for the final abundances are obtained 
by dividing these 1$\sigma $ errors by the squared root of the number of used 
lines. When only one line is available, a 0.1~dex value has been adopted. 

Most of the systematics arise from varying the adopted stellar parameters.
In order to properly quantify them, we generated a grid of test models with
varying the stellar parameters within $\pm$200~K in temperature (T$_{eff}$), $\pm$0.5
~dex in log-gravity (log~g) and $\pm$0.5~km~s$^{-1}$
in microturbulence velocity ($\xi$), and the abundances by 
$\pm$-0.1-0.2~dex accordingly, to reproduce the line depths. 
As a figure of merit of the statistical test we adopt
the difference between the model and the observed spectrum.
To quantify systematic discrepancies, this parameter is
more powerful than the classical $\chi ^2$ test, which is instead
equally sensitive to random and systematic scatters
\citep[see][ for more discussion and references therein]{ori04}.

All these alternative solutions turn out to be somewhat 
less statistically significant (typically at $1\le\sigma\le3$ level only) 
than our best-fit solution (P$>$90\%).
Moreover, since the
stellar features under consideration show a similar trend
with variation in the stellar parameters, although with different
sensitivity, {\it relative } abundances are less
dependent on stellar parameter assumptions,
reducing the systematic uncertainty
to $<$0.15~dex.

\section{Results and discussion}

For all of the 16 stars observed with CRIRES, Table~\ref{tab2} lists the measurement
of the radial velocities and abundances. 
The typical random error in the radial velocity estimate is $\approx$1 km/s.
Fifteen out of the 16 stars show heliocentric radial velocities consistent with 
being clustered around an average value of +57~km/s with 1$\sigma$ dispersion of 7~km/s.
The 15 stars with similar radial velocities also show very similar iron abundances, 
with an average [Fe/H]=-1.26$\pm$0.03 dex
and 1$\sigma$ dispersion of 0.10$\pm$0.02 dex.
For the same stars we also measured some $[\alpha$/Fe] abundance ratios, 
finding average values of 
[O/Fe]=+0.39$\pm$0.02\,dex ($\sigma$ = 0.09$\pm$0.02\,dex), [Mg/Fe] = +0.42$\pm$ 0.02\,dex 
($\sigma$ = 0.09$\pm$0.02\,dex), [Si/Fe] = +0.31$\pm$0.05\,dex ($\sigma$ = 0.15$\pm$0.03\,dex), and 
[Ti/Fe] = +0.15$\pm$0.04\,dex ($\sigma$ = 0.13$\pm$0.03\,dex).
Figure~\ref{alpha} shows the  measured $[\alpha$/Fe] abundance ratios as a 
function of [Fe/H].
These values of [$\alpha$/Fe] are fully consistent with those normally observed in the bulge stars: 
an enhancement of a factor of ${\sim}2\--3$ with respect to the solar value
 up to about solar metallicity.
We also measured carbon abundances, finding an average [C/Fe]=-0.18$\pm$0.02\,dex 
and 1$\sigma$ disperison of 0.08$\pm$0.02\,dex.
Some depletion of the [C/Fe] abundance ratio with respect to the solar value 
is normally observed in low-mass giant stars that are brighter than the RGB bump.

  \begin{figure*}
  \centering
  \includegraphics[width=8cm , height=8cm]{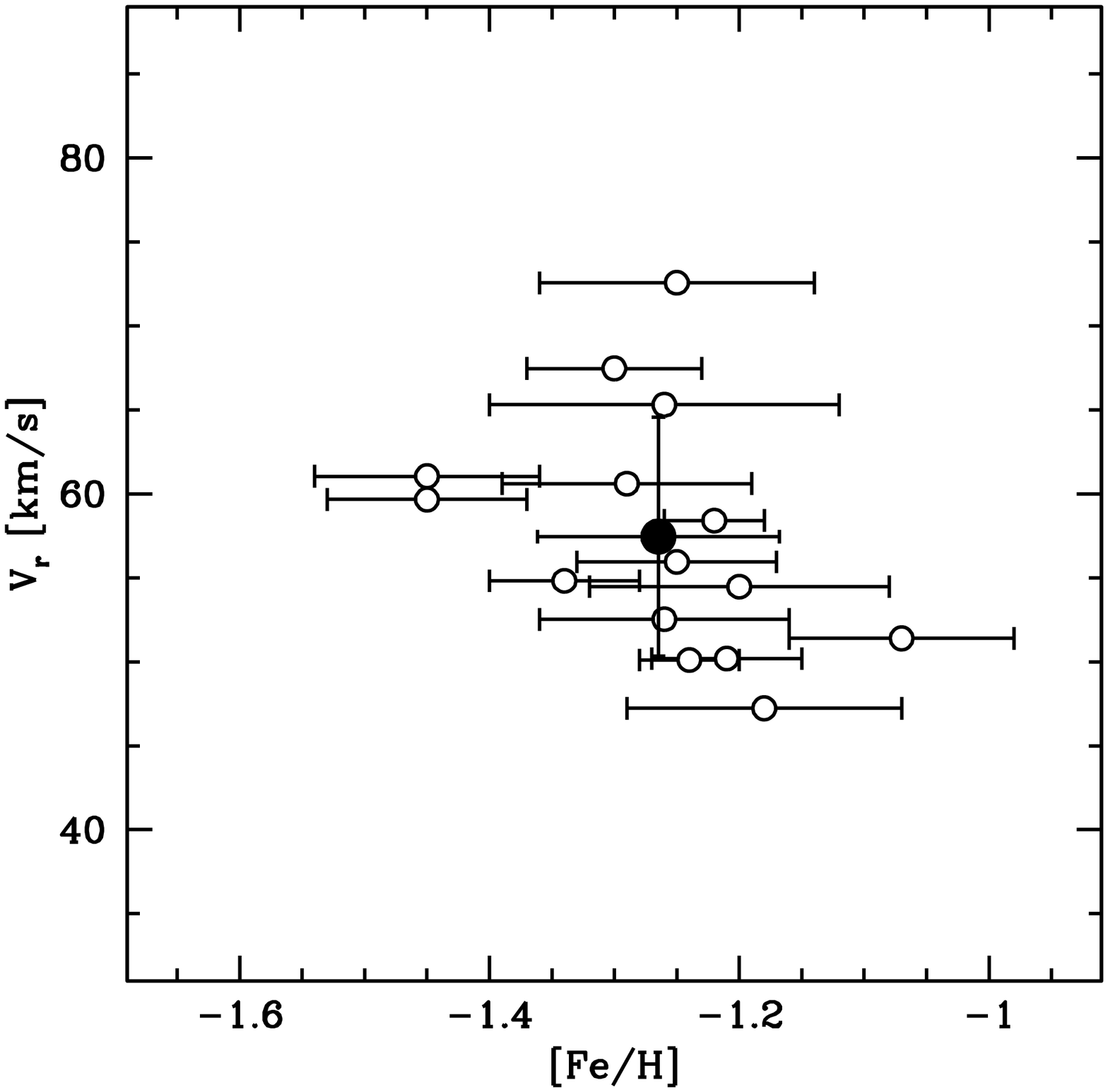}
  \includegraphics[width=8cm , height=8cm]{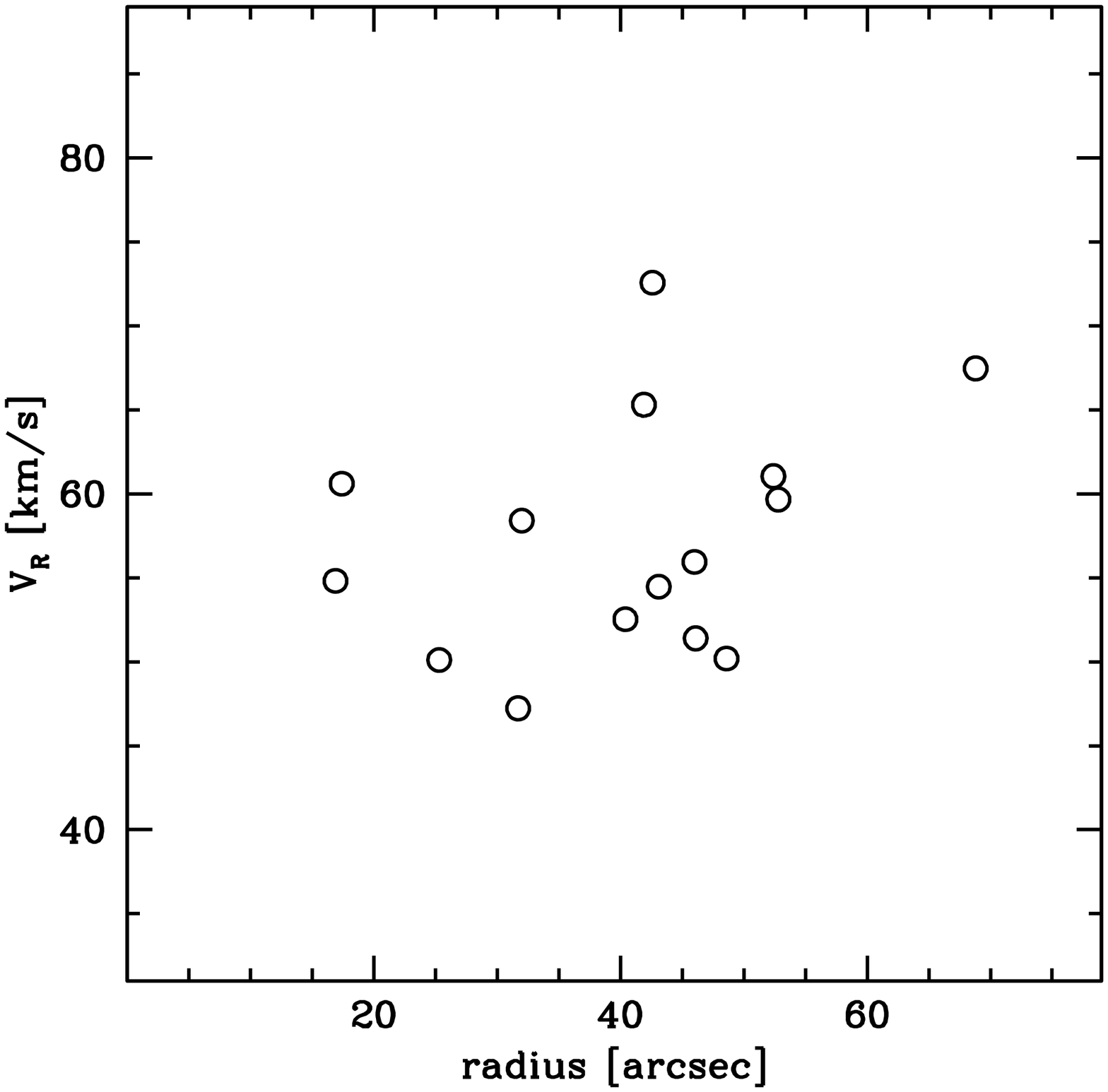}
   \caption{Heliocentric radial velocities for the 15 stars likely cluster members as a function of metallicity 
(left panel) and distance form the cluster center (right panel). 
The big dot in the left panel indicates the average value and 1$\sigma$ dispersion.}
              \label{rv}
    \end{figure*}

As shown in Fig.\ref{rv}, there is no evidence for any specific trend between the measured radial velocity and 
metallicity or distance from the cluster center, making all the 15 stars reasonable 
cluster member candidates.
Only one star (\#9) in our sample shows a radial velocity significantly different (by -104 km/s) 
from the average value of +57\,km/s shown by the 15 other stars, thus making it a candidate field star. 
Moreover, this star also has a metal content that is higher by about a factor of 3.

Eight stars in our sample (see Tab\,\ref{tab2}) have previously been observed by I02 
 at low resolution (R$ \leq$ 2000) in the optical range ($\lambda 4800\--6600 \AA$ and
 $\lambda 3980\--7848 \AA$). It is worth mentioning, however, that because the coordinates of
 such stars were not published by I02, the cross\--identification is based on visual inspection of
 their Fig.\,1, showing the position of the observed stars on a 2.5' $\times$ 2' image taken with
 the LNA 1.6\,m telescope. 
The membership of the eight stars in common based on derived radial velocities is consistent 
between the two studies with the only exception of star\,\#14 (ID\--I02:7), which according to I02 is a field star. 
However, in this case there might be a mismatch in the cross\--identification of the observed star 
owing the presence of a nearby second source, which is clearly resolved in our SofI image (see Fig.\,\ref{map-cmd}),
even better in the CRIRES acquisition camera, which has a resolution of 0.047"/pixel, but not as evident
in the optical image shown in Fig.\,1 of I02. 
Finally, because the position of star \#14 in the optical CMD of I02 is not shown, we cannot rule 
out the hypothesis that we actually observed different stars. In contrast to what can be seen
in our near\--IR map, in the optical image of I02 star \#14 and the
nearby source appear to have comparable magnitude. 
 
Nevertheless, what is most puzzling is the considerable disagreement on the derived radial velocities.
Based on 11 probable cluster members out of a total of 17 observed stars, I02 found a mean 
radial velocity of +114\, $\pm$14\,km/s, which is $\sim$60\,km/s higher than our estimate based on a
comparable sample. When we consider all the stars in common, but \#14, there is an average difference 
of 69\,km/s in the derived velocities and 1$\sigma$ dispersion of 19\,km/s. 
Our result is more consistent with a previous estimate based on integrated
spectroscopy \citep{arm-zinn88}, which reported a mean velocity of +35\,km/s although, admittedly, 
the integrated spectrum could have been heavily contaminated by
the presence of bright field sources in the center.

After carefully checking the derived wavelength calibration of CRIRES spectra (see \S\,\ref{obs})
we can rule out any systematic shift of $\sim$70\,km/s in radial velocity. 
On the other hand, it could be that at much lower resolution, a possible mismatch of spectral features 
produces a substantial shift in the radial velocity measurements.

Our result is instead in good agreement with Vasquez et al. (in prep., private communication), who
derive a mean radial velocity of +65\,km/s based on data obtained with FORS2 in multislit mode (R$\approx$2000).

Despite the large difference in terms of spectral resolution between this work and I02,
 the metallicity estimates for the stars in common are reasonably consistent, within $\sim$0.3\,dex, and $\le$0.1\,dex
when stars \#4 (ID\--I02: 5) and \#1 (ID\--I02: 13) are excluded. 
However, despite being listed among the candidate
cluster members, the metallicity tabulated by I02 for these two stars is 1\,dex higher than the mean 
metallicity of Terzan\,1 reported by the same authors. We found no obvious explaination for such a difference.


\begin{table*}
\caption{Chemical abundances of the stars toward Terzan~1 observed with CRIRES.}
\label{tab2}     
\centering                          
\begin{tabular}{ccccccccccc}  
\hline\hline
ID & T$\rm _{eff}$ & lg~g & V$_r$ & [Fe/H] & [C/Fe] & [O/Fe] & [Mg/Fe] & [Si/Fe] & [Ti/Fe]& ID-I02$^a$\\
\hline
1 &     4000 & 1.0 & +60  & -1.45$\pm$0.08 & -0.15$\pm$0.13 &  +0.36$\pm$0.11 & +0.44$\pm$0.10 & --	       & +0.03$\pm$0.13 &  13 \\						      
2 &     4000 & 1.0 & +61  & -1.45$\pm$0.09 & -0.15$\pm$0.13 &  +0.38$\pm$0.15 & +0.39$\pm$0.13 & --	       & +0.05$\pm$0.13 &  -- \\
3 &     4250 & 1.5 & +51  & -1.07$\pm$0.09 & -0.33$\pm$0.13 &  +0.28$\pm$0.15 & +0.39$\pm$0.12 & --	       & +0.14$\pm$0.13 &  12 \\
4 &     3800 & 1.0 & +67  & -1.30$\pm$0.07 & -0.10$\pm$0.12 &  +0.49$\pm$0.09 & +0.32$\pm$0.12 & +0.04$\pm$0.12& +0.21$\pm$0.12 &   5 \\
5 &     3800 & 1.0 & +53  & -1.34$\pm$0.06 & -0.06$\pm$0.12 &  +0.48$\pm$0.09 & +0.41$\pm$0.12 & +0.23$\pm$0.12& +0.03$\pm$0.12 &   9 \\	       
6 &     3800 & 1.0 & +65  & -1.26$\pm$0.14 & -0.14$\pm$0.17 &  +0.26$\pm$0.15 & +0.54$\pm$0.17 & +0.56$\pm$0.17& +0.03$\pm$0.17 &  -- \\	       
7 &     4000 & 1.0 & +56  & -1.25$\pm$0.08 & -0.15$\pm$0.13 &  +0.56$\pm$0.08 & +0.54$\pm$0.13 & --	       & +0.24$\pm$0.13 &   6 \\
8 &     3800 & 1.0 & +73  & -1.25$\pm$0.11 & -0.25$\pm$0.15 &  +0.41$\pm$0.12 & +0.45$\pm$0.16 & +0.45$\pm$0.15& +0.40$\pm$0.15 &  -- \\
9 &     3600 & 1.0 & -47  & -1.00$\pm$0.08 & -0.20$\pm$0.13 &  +0.39$\pm$0.09 & +0.31$\pm$0.18 & +0.05$\pm$0.13&             -- &  -- \\ 
10 &    4000 & 1.0 & +61  & -1.29$\pm$0.10 & -0.11$\pm$0.14 &  +0.46$\pm$0.11 & +0.55$\pm$0.14 & +0.37$\pm$0.14& +0.27$\pm$0.14 &  -- \\
11 &    4250 & 1.5 & +50  & -1.24$\pm$0.04 & -0.16$\pm$0.11 &  +0.46$\pm$0.07 & +0.49$\pm$0.11 & +0.36$\pm$0.11& +0.19$\pm$0.11 &  -- \\
12 &    3600 & 0.5 & +53  & -1.26$\pm$0.10 & -0.14$\pm$0.14 &  +0.41$\pm$0.11 & +0.33$\pm$0.15 & +0.23$\pm$0.14& +0.02$\pm$0.14 &   4 \\	     
13 &    3600 & 0.5 & +50  & -1.21$\pm$0.06 & -0.19$\pm$0.12 &  +0.26$\pm$0.10 & +0.25$\pm$0.12 & --	       &             -- &  -- \\	
14 &    3600 & 0.5 & +47  & -1.18$\pm$0.11 & -0.32$\pm$0.15 &  +0.33$\pm$0.13 & +0.39$\pm$0.15 & +0.20$\pm$0.15&             -- &  7  \\	     	  
15 &    3600 & 0.5 & +58  & -1.22$\pm$0.04 & -0.18$\pm$0.11 &  +0.38$\pm$0.05 & +0.46$\pm$0.11 & +0.33$\pm$0.11& +0.03$\pm$0.11 &  -- \\  
16 &    4000 & 1.0 & +54  & -1.20$\pm$0.12 & -0.20$\pm$0.16 &  +0.37$\pm$0.14 & +0.41$\pm$0.18 & +0.36$\pm$0.16& +0.32$\pm$0.16 &  11  \\ 
\hline\hline
\multicolumn{11}{l}{$^a$ Stars in common with I02 sample. The star reference identification is taken from their Table\,3.}
\end{tabular}
\end{table*}

\section{Conclusions}
We have obtained high\--resolution infrared spectroscopy of 16 giants in the heavily obscured
Galactic GC Terzan\,1.
Based on the derived radial velocity estimates, 15 out of 16 observed giants turned out to be candidate 
cluster members. We measured a cluster mean heliocentric radial velocity of +57\,km/s with 1 $\sigma$ dispersion of 7\,km/s.
This value is significantly less than what is reported in the \citet{harris} catalog based on the I02 study, but
it is in good agreement with the most recent low\--resolution optical estimate by Vasquez et al. (in prep.)

From our analysis of the candidate cluster members we find average
[Fe/H]= -1.26$\pm$0.03\,dex, [O/Fe]= +0.39$\pm$0.02\,dex, [Mg/Fe]= +0.42$\pm$0.02\,dex, 
[Si/Fe]= +0.31$\pm$0.05\,dex, and [Ti/Fe]= +0.15$\pm$0.04\,dex.
The derived metallicity estimates agree with the previous low\--resolution optical spectroscopic
study by I02, confirming that Terzan\,1, the innermost cluster in the bulge direction, is metal\--poor.

The [$\alpha$/Fe] enhancement measured in Terzan\,1 is
fully consistent with those measured in other bulge clusters observed in near\--IR by 
our group and in optical by other teams  \citep[see][and references therein]{val11}, as well as with the
estimate found in field giants located in the innermost bulge fields \citep{ric12}. 

The derived iron content measured on Terzan\,1 giants places this cluster on the metal\--poor tail of the observed
metallicity distribution function of the bulge field population 
\citep[see][and reference therein]{zoc08,oscar11,john11,john13,argos3}. Additionally, 
the observed $\alpha$\--enhancement makes Terzan\,1 chemically similar to the 
cluster and field bulge metal\--poor (i.e., [Fe/H]$\sim$1\,dex) component.

\begin{acknowledgements}

EV warmly thanks the telescope operators and night astronomers of Paranal working on UT1, for their 
competence and dedication while doing observations for this program.

RMR acknowledges support from grants AST\--1212095, AST\--1413755 from the National Science Foundation.

The authors thank Ivo Saviane and Sergio Velasquez for sharing information on their results on Terzan\,1
radial velocity measurements that will be presented in a forthcoming paper (Vasquez et al., in preparation).

NSO/Kitt Peak FTS data used here were produced by NSF/NOAO.

\end{acknowledgements}


\begin{thebibliography}{}

\bibitem[Alves\--Brito et al.(2006)]{albr06}
Alves\--Brito , A., Barbuy, B., Zoccali, M., et al. 2006, A\&A, 460, 269

\bibitem[Armandroff \& Zinn(1988)]{arm-zinn88}
Armandrfoff, T.\,E. \& Zinn, R.\, , 1988, \aj, 96, 92

\bibitem[Barbuy et al.(1999)]{bar99}
Barbuy, B., Renzini, A., Ortolani, S., et al., 1999,  A\&A, 341, 539

\bibitem[Barbuy et al.(2006)]{bar06}
Barbuy, B., Zoccali, M., Ortolani, S., Momany, Y., Minniti, D., Hill, V., Renzini, 
A., Rich, R. M., Bica, E., Pasquini, L., Yadav, R. K. S. 2006, \aap, 449, 349

\bibitem[Barbuy et al.(2007)]{bar07}
Barbuy, B., Zoccali, M., Ortolani, S., Minniti, D., Hill, V., Renzini, A., Bica, E., Gómez, A. 2007, \aj, 134, 1613

\bibitem[Barbuy et al.(2014)]{bar14}
Barbuy, B., Chiappini, C., Cantelli, E., et al. 2014, A\&A in press (arXiv1408.2438B)

\bibitem[Biemont \& Grevesse(1973)]{bie73}
Biemont, E., \& Grevesse, N. 1973, {\it Atomic Data and Nuclear Data Tables}, 12, 221 

\bibitem[Carretta et al.(2001)]{eug01}
Carretta, E., Cohen, J.~G., Gratton, R.~G., Behr, B. 2001, \aj, 122, 1469

\bibitem[Carretta et al.(2007)]{eug07}
Carretta, E., Bragaglia, A., Gratton, R.~G.et al, 2007, A\&A, 464, 967

\bibitem[Cohen et al. (1999)]{cohen99}
Cohen, J.~G., Gratton, R.~G., Behr, B. et al. 1999, \apj, 523, 739

\bibitem[Gonzalez et al.(2011)]{oscar11}
Gonzalez, O.,\,A., Rejkuba, M., Zoccali, M. et al. 2011, A\&A, 530, 54

\bibitem[Gratton et al.(2004)]{gra04}
Gratton, R. G., Sneden, C., \& Carretta, E. 2004, ARA\&A, 42, 385  

\bibitem[Gratton et al.(2006)]{gra06}
Gratton, R. G., Lucatello, S., Bragaglia, A., et al. 2006,  A\&A, 455, 271  

\bibitem[Grevesse \& Sauval(1998)]{gre98} 
Grevesse, N. \& Sauval, A.J. 1998, Space Sci. Rev., 85, 161

\bibitem[Idiart et al.(2002)]{idi02}
Idiart, T. P., Barbuy, B., Perrin, m. N., Ortolani, S., Bica, E., \& Renzini, A. 2002, \aap, 381, 472; I02

\bibitem[K\"{a}ufl et al. (2004)]{crires}
K\"{a}ufl, H.~U., Ballester, P., Biereichel, P. et al., 2004, SPIE,5492, 1218

\bibitem[Johnson et al(2011)]{john11}
Johnson, C.,I., Rich, R.\,M., Fullbright, J.\,P. et al. 2011, \apj, 732, 108

\bibitem[Johnson et al(2013)]{john13}
Johnson, C.,I., Rich, R.\,M., Kobayashi, C. et al. 2013, \apj, 765, 157

\bibitem[Johnson, Bernat \& Krupp(1980)]{jbk80}
Johnson, H. R., Bernat, A. P., \& Krupp, B. M. 1980, ApJS, 42, 501

\bibitem[Harris(1996)]{harris}
Harris, W.\,E. 1996, \aj, 112, 1487

\bibitem[McLean(1998)]{ml98}
McLean, I. et al. 1998, SPIE, 3354, 566

\bibitem[McWilliam(1997)]{mw97}
McWilliam, A. 1997, ARA\&A, 35, 503

\bibitem[Melendez \& Barbuy(1999)]{mel99}
Melendez, J., \& Barbuy, B. 1999, \apjs, 124, 527

\bibitem[Montegriffo et al.(1998)]{mon98}
Montegriffo, P., Ferraro, F.R., Fusi Pecci, F., \& Origlia, L.,
1995, \mnras, 276, 739

\bibitem[Ness et al.(2013)]{argos3}
Ness, M., Freeman, K., Athanassoula, E., et al. 2013, \mnras, 430, 836

\bibitem[Origlia, Moorwood \& Oliva(1993)]{ori93}
Origlia, L., Moorwood, A. F. M., \& Oliva, E. 1993, \aap, 280, 536

\bibitem[Origlia et al.(1997)]{ori97}
Origlia, L., Ferraro, F. R., Fusi Pecci, F., \& Oliva, E. 1997, \aap, 321, 859

\bibitem[Origlia, Rich \& Castro(2002)]{ori02}
Origlia, L., Rich, R. M., \& Castro, S. 2002, \aj, 123, 1559 

\bibitem[Origlia \& Rich(2004)]{ori04}
Origlia, L., \& Rich, R. M. 2004, AJ, 127, 3422 

\bibitem[Origlia, Valenti \& Rich(2005a)]{ori05a}
Origlia, L., Valenti, E., \& Rich, R. M. 2005b, \mnras, 356, 1276

\bibitem[Origlia et al.(2005b)]{ori05b}
Origlia, L., Valenti, E., Rich, R. M., \& Ferraro, F. R. 2005b, \mnras, 363, 897 

\bibitem[Origlia, Valenti \& Rich(2008)]{ori08}
Origlia, L., Valenti, E., \& Rich, R. M. 2008, \mnras, 388, 1419

\bibitem[Ortolani et al.(1999)]{ort99}
Ortolani, S., Barbuy, B., Bica, E., Renzini, A., Marconi, G., \& Gilmozzi, R., \aap, 350, 840

\bibitem[Rich, Origlia \& Valenti(2012)]{ric12}
Rich, R. M., Origlia, L., \& Valenti, E. 2012, \apj, 746, 59

\bibitem[Valenti, Ferraro \& Origlia(2010)]{val10}
Valenti, E., Ferraro, F.R., \& Origlia, L. 2010, \mnras, 402, 1729

\bibitem[Valenti, Origlia \& Rich(2011)]{val11}
Valenti, E., Origlia, L., \& Rich, R. M. 2011, \mnras, 414, 2690

\bibitem[Zoccali et al.(2004)]{zoc04}
Zoccali, M., Barbuy, B., Hill, V., et al. 2004, A\&A, 423, 507

\bibitem[Zoccali et al.(2008)]{zoc08}
Zoccali, M., Hill, V., Lecureur, A., et al. 2008, \aap, 486, 177

\end{thebibliography}
\end{document}